\documentclass[twocolumn,superscriptaddress,showpacs,amsfonts,amsmath,amssymb,prb]{revtex4}
\usepackage{graphicx}

\begin{document}

\title{Comparison between charge and spin transport in few layer graphene.}

\author{T. Maassen}
	\email{t.maassen@rug.nl} 
	\affiliation{Physics of Nanodevices, Zernike Institute for Advanced Materials, University of Groningen, Groningen, The Netherlands}
\author{F. K. Dejene}
	\affiliation{Physics of Nanodevices, Zernike Institute for Advanced Materials, University of Groningen, Groningen, The Netherlands}
\author{M. H. D. Guimar\~{a}es}
	\affiliation{Physics of Nanodevices, Zernike Institute for Advanced Materials, University of Groningen, Groningen, The Netherlands}
\author{C. J\'{o}zsa}
	\affiliation{Physics of Nanodevices, Zernike Institute for Advanced Materials, University of Groningen, Groningen, The Netherlands}
\author{B. J. van Wees}
	\affiliation{Physics of Nanodevices, Zernike Institute for Advanced Materials, University of Groningen, Groningen, The Netherlands}

\date{1 March 2011}

\begin{abstract}
Transport measurements on few layer graphene (FLG) are important because they interpolate between the properties of single layer graphene (SLG) as a true 2-dimensional material and the 3-dimensional bulk properties of graphite. In this article we present 4-probe local charge transport and non-local spin valve and spin precession measurements on lateral spin field-effect transistors (FET) on FLG. We study systematically the charge and spin transport properties depending on the number of layers and the electrical back gating of the device. We explain the charge transport measurements by taking the screening of scattering potentials into account and use the results to understand the spin data. The measured samples are between 3 and 20 layers thick, and we include in our analysis our earlier results of the measurements on SLG for comparison. In our room temperature spin transport measurements we manage to observe spin signals over distances up to $10~\mathrm{\mu m}$ and measure enhanced spin-relaxation times with an increasing number of layers, reaching $\tau_s\sim500~\mathrm{ps}$ as a maximum, about $4$ times higher than in SLG. The increase of $\tau_s$ can result from the screening of scattering potentials due to additional intrinsic charge carriers in FLG. We calculate the density of states (DOS) of FLG using a zone-folding scheme to determine the charge diffusion coefficient $D_C$ from the square resistance $R_S$. The resulting $D_C$ and the spin-diffusion coefficient $D_S$ show similar values and depend only weakly on the number of layers and gate induced charge carriers. We discuss the implications of this on the identification of the spin-relaxation mechanism.
\end{abstract}

 \maketitle
\section{\label{sec:Introduction}Introduction}
The electronic properties of exfoliated graphene have been studied in great detail \cite{RMP81_Castro2009, RMP82_Peres2010}, while the electron spin transport still brings up questions. The experimental spin-relaxation length $\lambda_S\sim2~\mathrm{\mu m}$ in single layer graphene (SLG) at room temperature is already promising \cite{N448_Tombros2007, PRB80_Popinciuc2009, PRB80_Jozsa2009, PRL105_Han2010} but is still at least one order of magnitude below theoretical predictions \cite{TEPJ148_Huertas-Hernando2007, PRL103_Huertas-Hernando2009}.
As the spin-relaxation is believed to be caused mainly by extrinsic scatterers in the substrate and on the surface of the graphene flake \cite{PRB80_Popinciuc2009, PRB80_Jozsa2009}, reducing the effect of these scatterers should lead to an improvement of the electronic transport \cite{NP5_Zhang2009} and an increase in $\lambda_S$. 

One way to avoid scatterers is separating the graphene flake from the substrate using suspended graphene, resulting in an increased charge carrier mobility $\mu$ \cite{SSC146_Bolotin2008, NN3_Du2008, c_Tombros2010}. Another way to enhance the transport properties is to screen the scattering potentials using few layer graphene (FLG). In a stack of graphene layers electrical potentials are screened by the outer layers with a screening length of about 1 to 5 layers \cite{PRB75_Guinea2007, PRB81_Koshino2010}, depending on the stacking order \cite{PRB81_Koshino2010}. This reduces the effect of external scatterers, resulting in only weakly influenced inner layers. The effect of screening of the gate induced charge carriers on the electrical transport in FLG has been observed in several groups' transport measurements \cite{NL9_Sui2009, APE1_Miyazaki2008}. While the spin transport in FLG was examined earlier \cite{JJAP46_Ohishi2007, APL90_Nishioka2007, APL92_Goto2008, SI7398_Han2009, PRB77_Wang2008, c_Yang2010}, there are no publications on the influence of screening on the spin transport properties. The influence of gate induced charge carriers on the spin signal is presented in Ref.~\citenum{APL92_Goto2008} while the effect of those charges on the spin-relaxation length still needs to be investigated. Along with studying the possible enhancement in spin transport using FLG, this investigation will also help with understanding the effect of possible multilayer inclusions in large scale graphene samples in future spintronic devices. 

This paper is organized as follows. In Sec.~\ref{sec:Sample_Fabrication} we describe the selection and preparation of FLG samples. In Sec.~\ref{sec:Charge_Transport} the charge transport properties of FLG are presented. We measure the dependence of the resistance on the number of layers and the gate induced charge carriers and explain the results considering electrical screening and a non-uniform background doping of the flake. At the end of the section we calculate the density of states (DOS) of FLG using a zone-folding scheme. In Sec.~\ref{sec:Spin_Transport} we discuss the spin transport properties of FLG as a function of the gate voltage ($V_g$), compare the results with SLG and show the evolution of the spin transport quantities as a function of the number of layers. Finally we compare the spin and charge transport and discuss the dominant spin-relaxation mechanism in our devices before the paper ends with the conclusions section.

\section{\label{sec:Sample_Fabrication}Sample Fabrication}

The presented measurements were performed at room temperature (RT) on mechanically exfoliated FLG flakes from highly oriented pyrolytic graphite (HOPG, from Advanced Ceramics, AB stacking) on a Si/SiO$_2$ substrate with an oxide thickness of $300~\mathrm{nm}$. We determine the thickness of the flakes using an atomic force microscope in tapping mode (TAFM). The measured thickness $t$ gives the number of layers by rounding down the quotient of $t$ and the spacing between two adjacent graphene layers $d_{SL}=0.335 ~ \mathrm{nm}$ (corresponding to the thickness of SLG), therefore the number of layers is $\left\lfloor t/d_{SL}\right\rfloor$. Due to the imprecise nature of the thickness measurements obtained with TAFM \cite{C46_Nemes-Incze2008} and due to the comparison between different thickness measurements on the same sample, we estimate an error in the number of layers for the FLG samples of about $1$ layer. Fig.~\ref{fig:FigI}(a) shows a scanning electron microscope (SEM) picture of a typical sample. The illustrated 20-layer FLG flake is contacted with several parallel aligned ferromagnetic cobalt electrodes obtained with electron beam lithography, e-beam evaporation of Co, and a standard lift-off technique. To avoid the conductivity mismatch and enhance the spin signal, we cover the graphene flake with an $0.8~ \mathrm{nm}$ thick insulating oxidized aluminum layer prior to the Co deposition, reaching contact resistances above $R_C=2 ~\mathrm{k \Omega}$. These contact resistances are larger than typical FLG resistances on a length scale of the spin-relaxation length $\lambda_S$, achieving in almost all cases non-invasive contacts \cite{PRB80_Popinciuc2009, APS57_Fabian2007}. The highly doped Si-substrate is contacted by an Au electrode for the electric gating of the device and controlling the amount of induced charge carriers $n_g$ in the system. The processing is given in detail in Ref.~\citenum{PRB80_Popinciuc2009}.

\section{\label{sec:Charge_Transport}Charge Transport}
\begin{figure}
\includegraphics[width=\columnwidth]{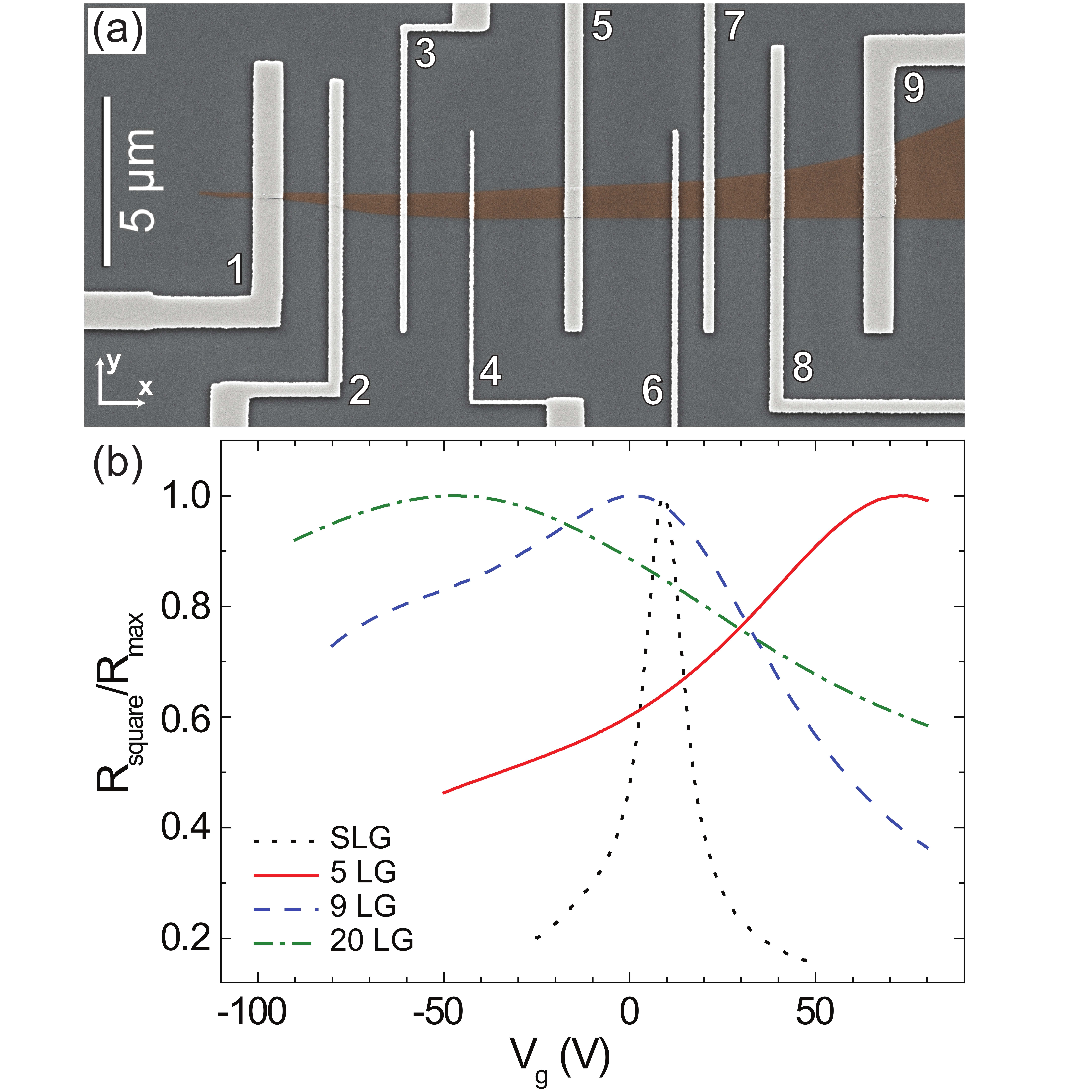} 
\caption{\label{fig:FigI}(Color online) (a) SEM image of a typical spin valve device. A 20-layer graphene flake contacted with several parallel ferromagnetic cobalt electrodes (light gray). (b) Normalized square resistance vs gate voltage for 1, 5, 9, and 20 layers.}
\end{figure}

The samples are first characterized by measuring the $V_g$ dependence of the square resistance $R_S$ of the FLG flake using local 4-probe geometry. Fig.~\ref{fig:FigI}(b) shows three typical measurements on FLG and one measurement on SLG. All curves show a maximum resistance (minimum conductivity $\sigma_\mathrm{min}$) at the respective $V_g = V_0$, marking the state where the Fermi energy $E_F$ coincides with the lowest DOS. $V_g=V_0$ is therefore the gate voltage with the lowest amount of induced charge carriers, corresponding to $n_g=0$. For our samples we get $n_g=\alpha (V_g-V_0)$ with $\alpha=7.2 \times 10^{10}~\mathrm{cm}^{-2}~\mathrm{V}^{-1}$, calculated using the SiO$_2$ thickness.

\begin{figure}
\includegraphics[width=\columnwidth]{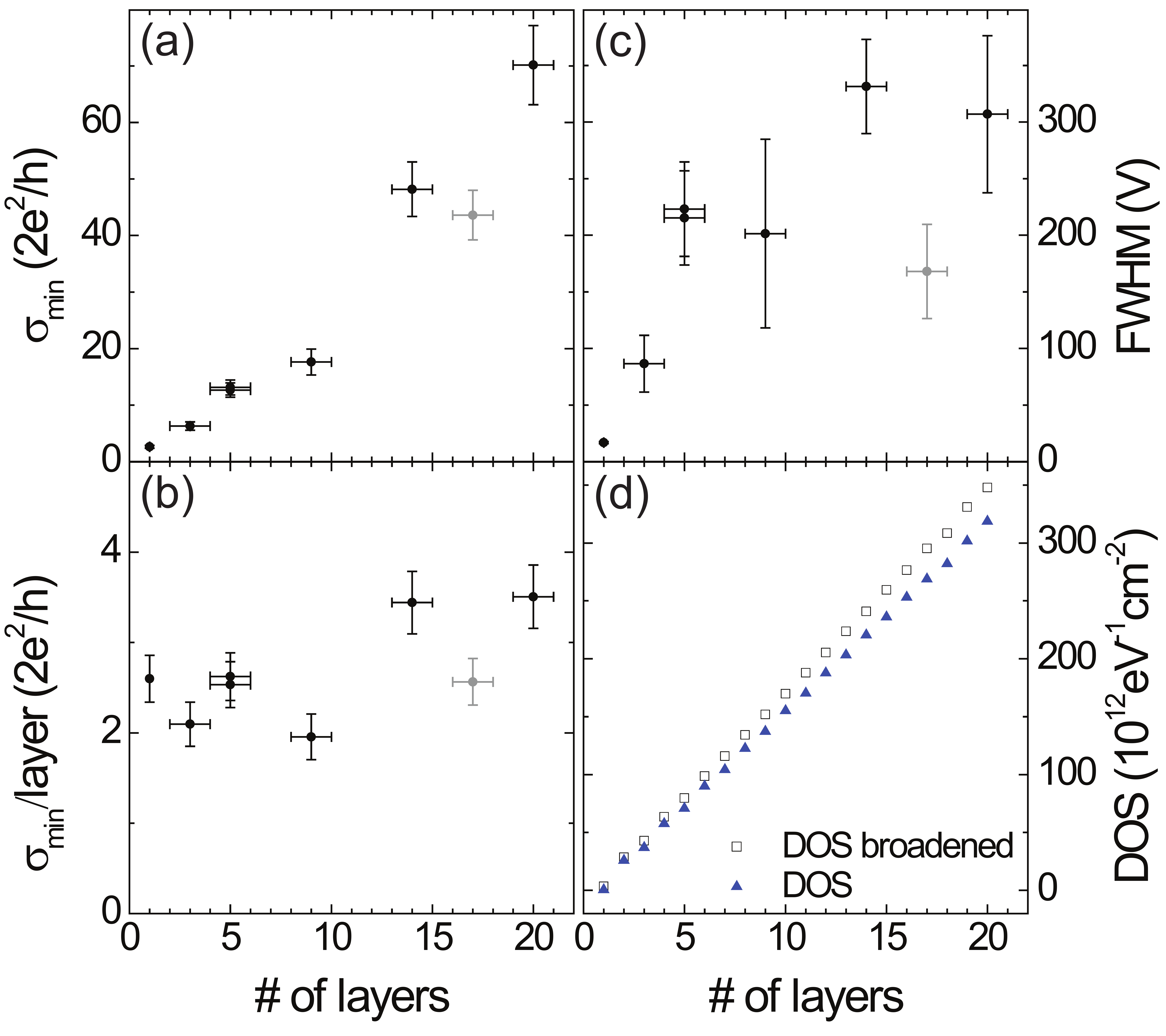} 
\caption{\label{fig:FigII}(Color online) (a) The minimum conductivity, (b) the minimum conductivity per layer and (c) the full width half maximum (FWHM) of the peak shaped resistance curves as a function of the number of layers. The gray points belong to a sample that showed overall unusual behavior \cite{foot_17layers}. (d) The DOS at $n_g=0$ as a function of the number of layers. The blue triangles show the calculated values using a zone-folding scheme, and the open black squares show the values including an energy broadening of $\mathrm{FWHM}\approx 60 ~\mathrm{meV}$.}
\end{figure}

The values of $\sigma_\mathrm{min}$ are displayed in Fig.~\ref{fig:FigII}(a) as a function of the number of layers. The conductivity increases approximately linearly with increasing thickness. This can also be seen by the fairly constant value for $\sigma_\mathrm{min}$ per layer plotted in Fig.~\ref{fig:FigII}(b) and can be explained by the linear increase of the DOS with the number of layers as presented in Fig.~\ref{fig:FigII}(d). The DOS was calculated using a zone-folding scheme as described at the end of this section and the displayed points show the DOS at $n_g=0$, corresponding to the minimum value of the DOS. The DOS is plotted with and without taking into account energy broadening resulting in slightly different slopes. The linear increase of the DOS points to a weak influence of the graphene layer stacking on the DOS per layer. For thicker samples we see a small increase of $\sigma_\mathrm{min} / \mathrm{layer}$ \cite{foot_17layers}. The conductance per layer of 20-layer graphene of $\sigma_\mathrm{min}/\mathrm{layer}\sim3.5 \times 2e^2/h$ increases further to $\sigma/\mathrm{layer}\sim8.5 \times 2e^2/h$ in bulk graphite \cite{foot_sigmamin,PRB41_Matsubara1990}. This rise in $\sigma_\mathrm{min} / \mathrm{layer}$ for thicker samples could be explained by a stronger influence of the coupling between the layers with increasing thickness, which is consistent with the interlayer coupling tight binding parameter rising from bilayer graphene (BLG) to graphite \cite{PRB76_Malard2007}. 

\begin{figure}
\includegraphics[width=\columnwidth]{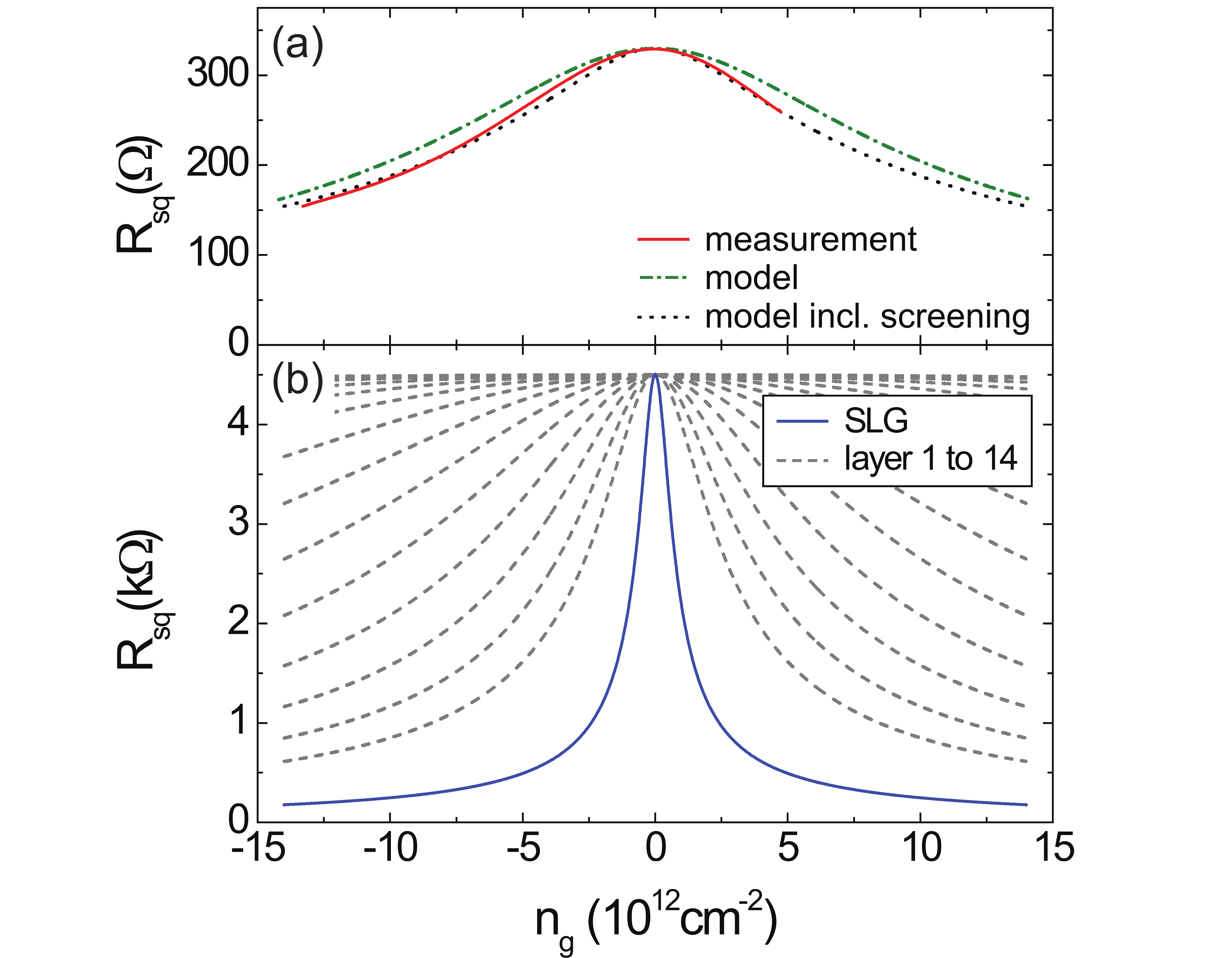} 
\caption{\label{fig:FigIII}(Color online) (a) $R_S$ vs $V_g$ for a 14-layer thick sample from a 4-probe measurement (red, solid) and from our model excluding (green, dash-dotted) and including the screening of the gate voltage (black, dotted). (b) Sketch of the model for a FLG stack. The blue solid curve shows the modeled SLG resistance $R_S^\mathrm{model}$ considering $\mu = 2520~\mathrm{cm^2V^{-1}s^{-1}}$ and a minimum carrier density of $n_0=0.55 \times 10^{12}~\mathrm{cm^{-2}}$. The gray dashed curves show the square resistances for the 14-layers that, when combined, result in the black dotted curve for the modeled stack resistance in (a).}
\end{figure}

While $\sigma_\mathrm{min}$ increases with the number of layers, the influence of the gate voltage on the resistance is reduced with increasing thickness. This can be seen by the increased full width half maximum (FWHM) of the peak shaped resistance curve (see Fig.~\ref{fig:FigI}(b) and Fig.~\ref{fig:FigII}(c)) and can be explained by the distribution of the induced charges over the layers. The red, solid curve in Fig.~\ref{fig:FigIII}(a) shows $R_S$ vs $V_g$ measured on a 14-layer sample. Assuming an equal division of the charges between the SLG-like layers, we get a broadened resistance curve (see Fig.~\ref{fig:FigIII}(a), green, dash-dotted curve). Taking the screening of the extrinsic potentials (including the gate voltage) by a few layers\cite{PRB75_Guinea2007} into account, we get even better agreement between the modeled resistance curve and the measured one (see Fig.~\ref{fig:FigIII}(a), black, dotted curve). The influence of $V_g$ on $R_S$ can be described following an easy resistor model (see Fig.~4(a)~in Ref.~\citenum{NL9_Sui2009}). The FLG flake is modeled as parallel resistors (the graphene sheets) contacted via an interlayer resistance $R_{int}=\rho_c~d_{SL}/A$ at the source and drain, where $\rho_c \approx 0.1 ~ \mathrm{\Omega cm}$ is the conductance along the c-axis of HOPG \cite{PRB41_Matsubara1990} and $A$ is the contact area. With our contact areas of $A\sim0.5\times0.5 ~ \mathrm{\mu m^2}$, we get $R_{int}\sim1 ~ \mathrm\Omega$, much smaller than typical SLG resistances, $R_S\sim2 ~ \mathrm{k\Omega}$. No further conductance between the layers is considered. 

Using Thomas-Fermi screening \cite{PRB81_Koshino2010}, the total induced charge is screened approximately exponentially and distributed over the layers \cite{PRB75_Guinea2007, NL9_Sui2009} shown in Fig.~\ref{fig:FigIII}(b). We assume the resistance per layer to be equal to a modeled SLG resistance $R_S^\mathrm{model} = (e\mu\sqrt{{n_\mathrm{ind}}^2+{n_{0}}^2})^{-1}$, where $e$ is the electron charge and $n_\mathrm{ind}$ and $n_{0}$ are the induced and the minimum charge carrier densities per layer. The total induced charge carrier density $n_g$ is distributed over the layers, with an exponential decay of the induced charge carrier density per layer $n_\mathrm{ind}^i$ (compare Ref.~\citenum{NL9_Sui2009}, eq. (5)). Fig.~\ref{fig:FigIII}(b) shows $R_S^\mathrm{model}$ for SLG with $\mu = 2520~\mathrm{cm^2V^{-1}s^{-1}}$ and $n_0=0.55 \times 10^{12}~\mathrm{cm^{-2}}$ (blue, solid) and the calculated $R_S$ for the different layers in the stack (gray, dashed) as a function of $n_g$. Considering the screening length, we follow Ref.~\citenum{PRB75_Guinea2007} with $\lambda=3 - 5~\mathrm{layers}$ and use the best fit to our data, $\lambda=3~\mathrm{layers}$. In the case of SLG it is $n_\mathrm{ind}=n_g$ and for FLG $\sum{n_\mathrm{ind}^i}=n_g$. The modeled layers closest to the gate still experience a strong resistance change by changing $n_g$, while the resistance of layers farther away is almost unaffected. The resulting resistance of the 14-layer stack is plotted in Fig.~\ref{fig:FigIII}(a) (black, dotted) together with the measured $R_S$ of the 14-layer graphene sample (red, solid). We see good agreement between the two curves. In the case of modeling the resistance excluding the screening we have to investigate $\lambda~\rightarrow~\infty$ and get as a result the green dash-dotted curve in Fig.~\ref{fig:FigIII}(a), which does not fit as well to our measurements as the one that includes screening. 

\begin{figure}
\includegraphics[width=\columnwidth]{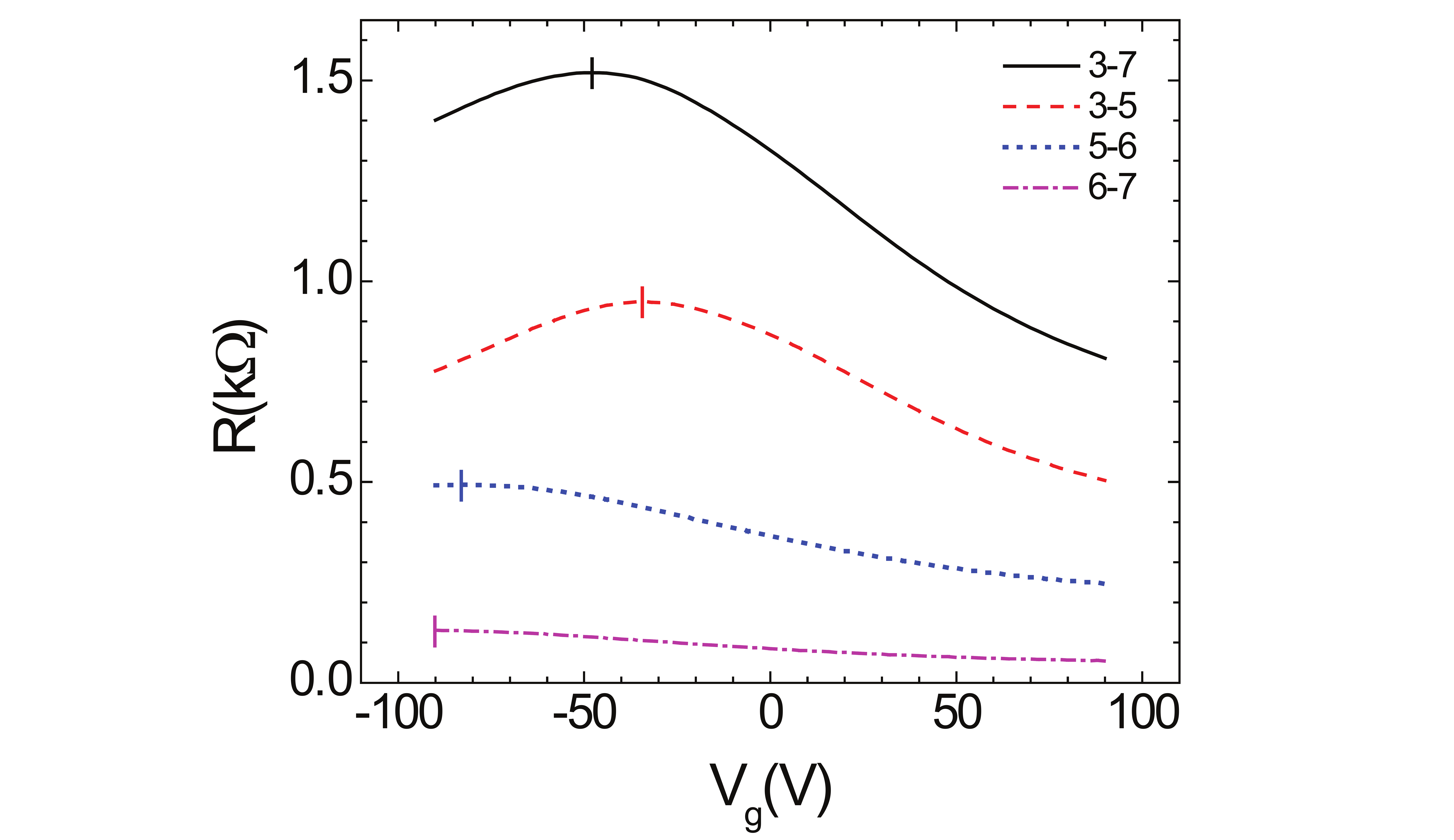} 
\caption{\label{fig:FigIV}(Color online) 4-probe measurements of the local resistance of the 20-layer graphene sample shown in Fig.~\ref{fig:FigI}(a) as a function of $V_g$. The black solid curve shows the resistance on a $9~\mathrm{\mu m}$ long strip (between contact 3 and 7); the other curves represent the resistances on parts of this strip, between contacts 3 and 5 ($5~\mathrm{\mu m}$, red, dashed), contacts 5 and 6 ($3~\mathrm{\mu m}$, blue, dotted), and contacts 6 and 7 ($1~\mathrm{\mu m}$, magenta, dash-dotted). The vertical lines show the position of the respective maximum resistance $V_0$.}
\end{figure}

While the shape of the resistance curve for, e.g., the 14-layer graphene sample can be easily modeled, some samples show a further broadened or an asymmetric resistance curve as a function of $V_g$ (see, e.g., the curve for 9-layer graphene in Fig.~\ref{fig:FigI}(b)). This can be explained as follows: Fig.~\ref{fig:FigIV} shows the resistance of a 20-layer graphene sample, measured on different parts of the FLG flake. The black solid curve was measured on a $9~\mathrm{\mu m}$ long strip, while the other three curves represent the resistance of sections of this strip. Between the three sections we see a shift of $V_0$ by $\sim50 ~\mathrm{V}$. Adding up the resistances, results for the full distance in the broadened curve (black, solid) with a lower maximum resistance compared to a sample with a fixed position of $V_0$ for all sections. The shift of $V_0$ is caused by a non-uniform background doping of the flake that could be due to a locally different doped substrate, resist residues on the surface of the flake, or the metal contacts. This effect could explain the asymmetric shape of the resistance curves and the spread in the values for the minimum conductivity per layer and for the FWHM in Fig.~\ref{fig:FigII}(b) and (c), respectively. The same effect has been observed in SLG \cite{SSC149_Blake2009}. 

%
In the following section we will discuss the spin transport properties and compare spin with charge diffusion in FLG to discuss the spin-relaxation mechanism. Therefore, we need to calculate the charge diffusion coefficient $D_C$ based on $R_S$ using the Einstein relation 
\begin{equation}
\sigma = e \nu D_C, 
\label{eq:einstein}
\end{equation}
where $\nu(E)$ is the energy dependent DOS and $\sigma(E)=1/R_S$. In order to easily calculate the DOS for FLG we use the fact that the tight-binding Hamiltonian of a FLG graphene system can be, in a good approximation, separated into sets of BLG-like and SLG-like Hamiltonians \cite{PRB76_Koshino2007}. This approach was already experimentally validated by infrared spectroscopy \cite{PRL102_Orlita2009,PNAS107_Mak2010}.

To determine the number and shape of the BLG- and SLG-like bands we apply a zone-folding scheme introduced by Mak \textit{et. al} \cite{PNAS107_Mak2010} that reduces the 3-dimensional (3D) band structure of graphite into a 2-dimensional (2D) band structure for FLG. This approach uses the fact that the confinement in the $z$-direction (perpendicular to the FLG flake) induces standing waves and therefore a quantization on the wave-vector $k_z$. The quantization can be represented by cutting planes in the 3D Brillouin zone (BZ) of graphite that cut through different regions of the BZ due to the different symmetry-groups for FLG with even or odd number of layers \cite{PRB79_Malard2009}. For an odd number of layers, there is always a cutting plane through the H point, which introduces a linear dispersion band similar to SLG. For an even number of layers, such a cutting plane is not present. The other cutting planes (if any) do not pass through the borders of the graphite BZ and introduce BLG-like bands with different effective masses.

To define the energy dispersion for graphite, we use a simple tight-binding approach consisting of only the hopping parameter for next-neighbors, $\gamma_0 = 3.15 \ eV$, and an interlayer coupling of $\gamma_1 = 0.37 \ eV$. The inclusion of other interlayer and intralayer coupling parameters have a minor effect on our results since they are smeared out by a broadening introduced by temperature, impurities, and other disorder potentials \cite{PNAS107_Mak2010}. To account for such effects, we include a Gaussian broadening in the DOS in the same way as shown in Ref.~\citenum{PRB80_Jozsa2009}. The resulting DOS at $n_g=0$ increases linearly with the number of layers and is presented in Fig.~\ref{fig:FigII}(d). In the figure the calculated values using only the zone-folding scheme are presented along with the values including an energy broadening of $\mathrm{FWHM}\approx 60 ~\mathrm{meV}$. This broadening also takes into account the effect of the electron and hole pockets around $E_F$ present in graphite \cite{PRB74_Partoens2006}.

\section{\label{sec:Spin_Transport}Spin Transport}
\begin{figure}
\includegraphics[width=\columnwidth]{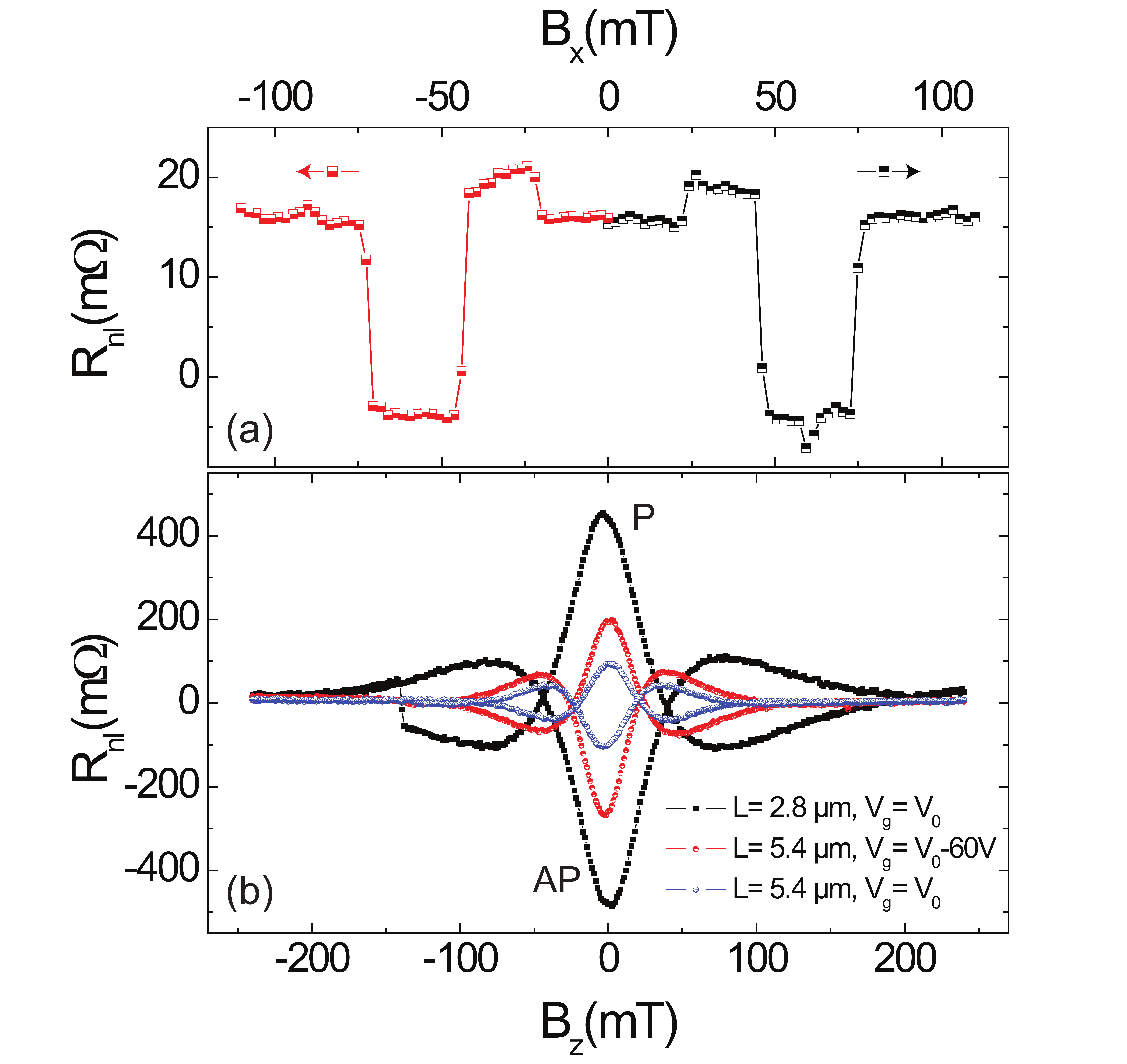} 
\caption{\label{fig:Fig1}(Color online) (a) Non-local spin valve signal of a 7-layer graphene sample. The sweep directions of the magnetic field are indicated
(red and black arrows). The distance between the inner electrodes is $L = 8~\mathrm{\mu m}$. (b) Hanle precession measurements of a 5-layer graphene sample at the gate voltage resulting in the minimum conductivity $V_g=V_0$ for $L=2.8~\mathrm{\mu m}$ (black, largest amplitude) and $L=5.4~\mathrm{\mu m}$ (blue, smallest amplitude) and in the hole doped state at $V_g=V_0-60~\mathrm{V}$ for $L=5.4~\mathrm{\mu m}$ (red, intermediate amplitude). The precession is measured for the parallel (P) and antiparallel (AP) configuration of the inner contacts. The curve for $L=2.8~\mathrm{\mu m}$ shows a switch from the P to the AP state at $-140~\mathrm{mT}$. This is due to the fact that at relatively high fields the non-avoidable in-plane component of the perpendicular field switches the magnetization of one of the inner electrodes.}
\end{figure}

Now we examine the spin transport properties of FLG. Fig.~\ref{fig:Fig1}(a) shows a typical non-local spin valve measurement \cite{N448_Tombros2007} on FLG. Sending a current $I$ from electrode 5 to electrode 1 (see Fig.~\ref{fig:FigI}(a)) generates a spin accumulation at electrode 5. The spins diffuse on both sides of the electrode along the flake and generate a voltage drop $V_{nl}$ between electrodes 6 and 9, defining the non-local resistance $R_{nl}=V_{nl}/I$. Switching the magnetization of one of the inner electrodes (5 or 6) using an in-plane magnetic field results in a sign change of $R_{nl}$ (see Fig.~\ref{fig:Fig1}(a)). When the outer contacts (1 or 9) are located within the spin-relaxation length, additional switches can be observed \cite{N448_Tombros2007}. The spin valve measurement in Fig.~\ref{fig:Fig1}(a) is taken on a 7-layer graphene sample with an inner contact distance of $L=8 ~ \mathrm{\mu m}$. Including the additional switch at small field values, we see a spin signal over a distance of $L=10 ~ \mathrm{\mu m}$. It is worth noting that this is the longest distance over which a spin signal has ever been reported for graphene based devices.

For further analysis of the spin transport we perform Hanle spin precession measurements \cite{APS57_Fabian2007}. They are performed in the same geometry as the spin valve measurements with the magnetic field $\vec{B}$ pointing now perpendicular to the sample plane causing the injected, in-plane oriented spins to precess. The spin dynamics are described by the Bloch equation for the spin accumulation $\vec{\mu_S}$: \cite{APS57_Fabian2007}
\begin{equation}
D_S\mathbf\nabla^2 \vec{\mu_S} - \frac{\vec{\mu_S}}{\tau_S} + \vec{\omega_0}\times\vec{\mu_S}=\vec{0}.
\label{eq:Bloch}
\end{equation}
The first term on the left-hand side describes the spin-diffusion represented by the spin-diffusion coefficient $D_S$, and the second term describes the spin-relaxation with the spin-relaxation time $\tau_S$. The third term describes the precession with the Larmor frequency $\vec{\omega_0}=g\mu_B/\hbar \ \vec{B}$, where $g=2$ is the effective Land\'{e} factor and $\mu_B$ is the Bohr magneton. In Fig.~\ref{fig:Fig1}(b) three Hanle measurements on a 5-layer graphene sample are presented. Each curve consists of the non-local signal acquired for the parallel (P) and antiparallel (AP) orientation of the inner contacts. The black and the blue dots represent the measurements for $L=2.8~ \mathrm{\mu m}$ and $L=5.4~ \mathrm{\mu m}$, respectively, at the gate voltage $V_0$. The red curve is measured on the longer distance at $V_g= V_0-60~\mathrm{V}$, where electron charges are induced by the gate. The amplitude for the measurement with increased $L$ is smaller due to additional spin-relaxation, as the spins have to travel a longer distance resulting in a longer time interval for spin-relaxation. In addition to the change in the amplitude, a shift in the $B$-field values for the crossing points of the parallel and the antiparallel precession curve is visible for the two curves measured at $V_g=V_0$. The crossing points represent the $B$-field value where the spins have, on average, precessed for $90^\circ$, resulting in both configurations in a signal of $R_{nl} \approx 0$. An increased distance $L$, corresponding to an increased travel time for the spins, therefore decreases the $B$-field which results in $90^\circ$-precession \cite{PRB80_Popinciuc2009, APS57_Fabian2007}. 

The measurements show that the spin signal can be enhanced by inducing more charge carriers (see enhanced spin signal comparing the measurement at $V_g= V_0-60~\mathrm{V}$ and $V_g=V_0$). This was also observed for SLG \cite{PRB80_Jozsa2009}. 

\begin{figure}
\includegraphics[width=\columnwidth]{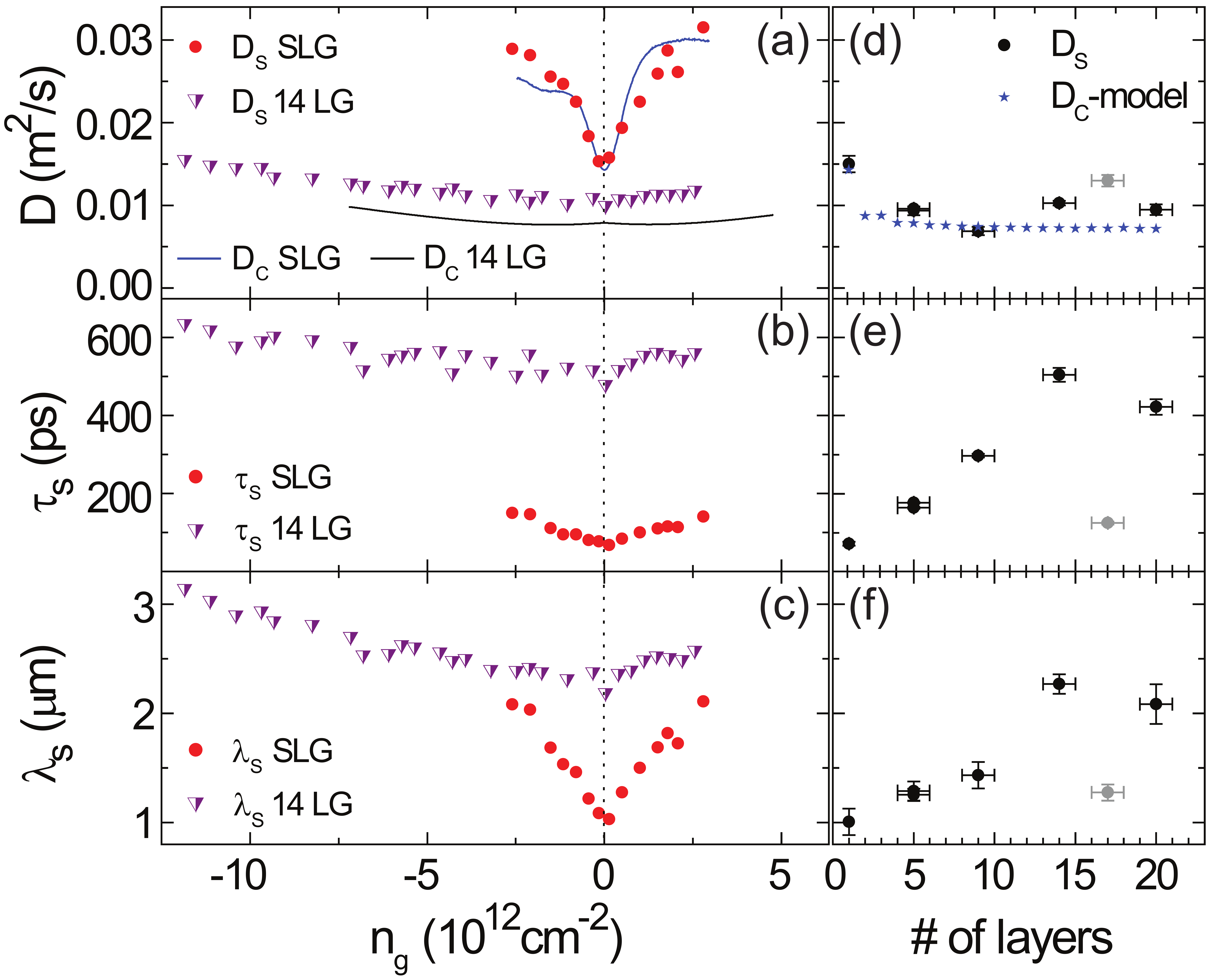} 
\caption{\label{fig:Fig3}(Color online) $D_S$, $\tau_S$ and $\lambda_S$ (a)-(c) as a function of the induced charge carriers $n_g$ for 14-layer graphene and SLG and (d)-(f) as a function of the number of layers at $V_g~=~V_0$. The gray points belong to a sample that showed overall unusual behavior \cite{foot_17layers}. In addition to $D_S$, (a) also shows $D_C$ for 14-layer graphene and SLG and (d)~shows the modeled $D_C$ at $V_g~=~V_0$ assuming $R_S=5~\mathrm{k \Omega/layer}$.}
\end{figure}

The Hanle curves can be fitted with the solutions of the Bloch-equation~(\ref{eq:Bloch}), yielding the spin transport quantities $D_S$ and $\tau_S$. Those solutions are calculated with the injector considered to be a spin current source and the detector considered to be a non-invasive spin voltage probe \cite{APS57_Fabian2007, PRB80_Popinciuc2009}. To exclude (small) spurious background effects, we subtract the AP from the P curve and fit the result. For several FLG samples a set of precession measurements was performed for different induced charge-carrier densities $n_g$. The spin transport quantities $D_S$ and $\tau_S$ and the spin-relaxation length $\lambda_S=\sqrt{D_S \tau_S}$ are plotted as a function of $n_g$ in Fig.~\ref{fig:Fig3}(a), (b), and (c), respectively. Here the results for SLG from Ref.~\citenum{PRB80_Jozsa2009} are compared with a 14-layer graphene sample representing the results for FLG. The general dependence of the quantities is the same for all samples. All curves show a minimum at $n_g=0$ (corresponding to $V_g~=~V_0$). The change in the three different quantities as a function of $n_g$ is minor compared to SLG. This can be explained by the fact that in FLG far more intrinsic charge carriers are present due to the changed band structure compared to SLG, masking the effect of the induced charge carriers $n_g$ (see Sec. \ref{sec:Charge_Transport}). 

Fig.~\ref{fig:Fig3} (d), (e), and (f) show $D_S$, $\tau_S$, and $\lambda_S$, respectively, at $V_g=V_0$ as a function of the number of layers. Besides a drop from SLG to BLG, $D_S$ shows no identifiable dependence and is approximately constant, while $\tau_S$ increases linearly with the number of layers until this trend is reduced for the thickest samples, reaching a maximum of $\tau_S\sim500~\mathrm{ps}$. With this result, $\tau_S$ still stays far below the spin-relaxation times of $20~\mathrm{ns}$ measured at RT with electron spin resonance (ESR) in bulk graphite \cite{PR118_Wagoner1960} or $\tau_S=55~\mathrm{ns}$ in graphene at $T=150~\mathrm{K}$ \cite{PSSb246_Ciric2009}. There is, at this time, no explanation for that difference. The linear increase of the measured $\tau_S$ as a function of the number of layers can be explained by the expected screening of scattering potentials due to the linear increase of intrinsic charge carriers as a function of the number of layers (compare Fig.~\ref{fig:FigII}(d)). 

The constant value for $D_S$ for more than one layer shows that the change in the band structure and the screening does not have a strong influence on the spin-diffusion. On the other hand the spin-relaxation length increases with the number of layers and is doubled at $V_g=V_0$ between 1 and 20 layers. The effect of the induced charge carriers on the spin transport quantities is weak for thicker samples. Therefore, we see only a small increase of $\lambda_S$ in FLG for $n_g \neq 0$ compared to SLG (Fig.~\ref{fig:Fig3}(c)) and reach a maximum value of $\lambda_S\sim3 ~ \mathrm{\mu m}$. This is also because outer scattering potentials are already screened by the intrinsic charge carriers in the thicker samples. 

Our results do not show values for $\lambda_S$ as high as reported by Goto \textit{et al.} \cite{APL92_Goto2008} ($\lambda_S \gg 8 \mathrm{\mu m}$). This is probably due to the fact that the reported values were derived indirectly from spin valve measurements on short distances of $L\sim300~\mathrm{nm}\ll 8 \mathrm{\mu m}$, making it difficult to conclude the behavior over long distances. We also note that spin valve measurements are, in general, less conclusive for spin transport properties than Hanle precession measurements \cite{APS57_Fabian2007}. 

Han \textit{et al.} discuss in Ref.~\citenum{PRL105_Han2010} different behaviors of the spin signal depending on the induced charge carriers for different kinds of contact interfaces. We believe that in our samples the interface mainly affects the polarization of the injected current and has only weak influence on spin-relaxation and spin-diffusion. While we have seen reduced spin signals for low $R_C$ in SLG samples before \cite{PRB80_Popinciuc2009}, the measurements presented here were performed on samples with $R_C$ values that lead to an $R$ parameter of $R\ge0.1~\mathrm{\mu m}$. As described in Ref.~\citenum{PRB80_Popinciuc2009}, $R/\lambda_S$ represents the ratio between the contact resistance and the graphene resistance over one spin-relaxation length. The values that we found for $R$ show that the contacts in our samples are non-invasive. Hence, we can rule out effects of spins escaping into the cobalt electrodes, fringe fields, and interface spin scattering. This is also supported by the fact that we have observed spin transport under electrically floating cobalt electrodes without any measurable effects on the spin signal in SLG and FLG.

In addition to $D_S$, Fig.~\ref{fig:Fig3}(a) and (d) show the charge diffusion coefficient $D_C$, which is calculated using equation~(\ref{eq:einstein}) requiring the DOS. For FLG we use the DOS obtained by the zone-folding scheme as discussed in Sec.~\ref{sec:Charge_Transport}, assuming a broadening of $\mathrm{FWHM}\approx 60 ~\mathrm{meV}$~\cite{foot_60meV} and, in the case of SLG, the broadened DOS discussed in Ref.~\citenum{PRB80_Jozsa2009}. In Fig.~\ref{fig:Fig3}(a) we use the measured values for the conductivity of the samples as a function of $V_g$, while in Fig.~\ref{fig:Fig3}(d) we assume a fixed resistance per layer of $R_S=5~\mathrm{k \Omega/layer}$ at $V_0$ (compare Fig.~\ref{fig:FigII}(b)). For SLG the two diffusion coefficients have very similar values independent of $n_g$. In 14-layer graphene $D_C$ is $\sim20 \%$ smaller than $D_S$ (and $\sim50\%$ smaller than in SLG), while both coefficients show only a slight change as a function of $n_g$ ($D_C \propto D_S$, see Fig.~\ref{fig:Fig3}(a)). As a function of the number of layers, $D_C$ and $D_S$ are approximately constant after the values drop between SLG and BLG by roughly $\sim50\%$. We still see a slight decrease from $2$ to $5$ layers. This shows that here $D_C$ behaves the same way as $D_S$ and is not affected by the changing band structure or the screening. Similar to the case for SLG, the Coulomb electron-electron interactions still play a minor role in the scattering \cite{PRB80_Jozsa2009}. The main factor limiting diffusion is still impurity potential scattering.

\begin{figure}
\includegraphics[width=\columnwidth]{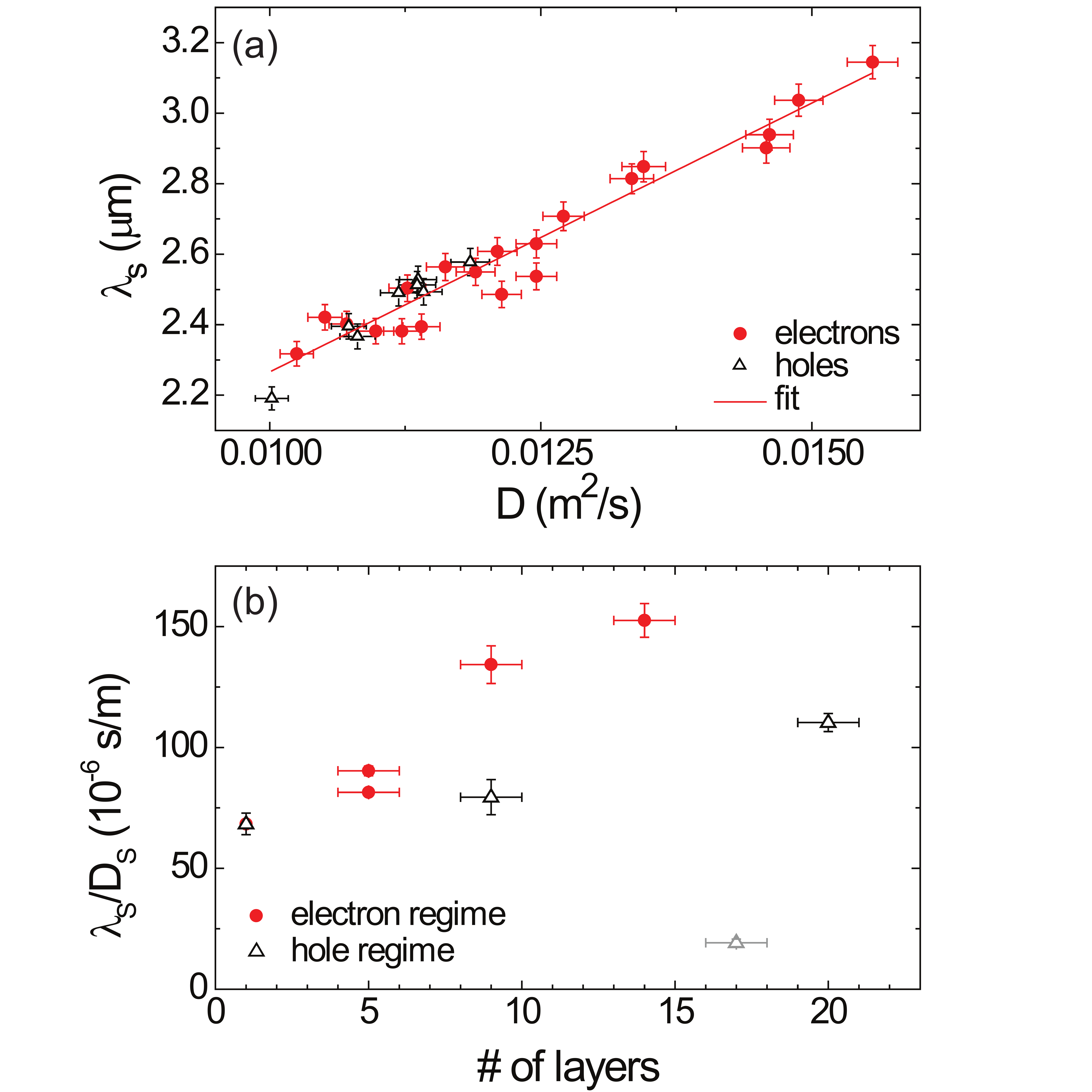} 
\caption{\label{fig:Fig4}(Color online) (a) Linear relationship between $\lambda_S$ and $D_S$, extracted from Fig.~\ref{fig:Fig3}(a) and (c). The linear fit is performed on the red points taken in the electron conduction regime, and the black open triangles were taken in the hole regime. (b) Slopes of the linear fits of $\lambda_S$ vs $D_S$ for different samples as a function of the number of layers. The increase in the slopes is different for the electron and hole conduction regimes. The gray point belongs to a sample that showed overall unusual behavior \cite{foot_17layers}.}
\end{figure}

This can also be seen when we plot $\lambda_S$ as a function of the corresponding $D_S$. We observe a linear dependence for the electron and the hole conduction regimes. Fig.~\ref{fig:Fig4}(a) shows this plot for the 14-layer graphene sample (data extracted from Fig.~\ref{fig:Fig3}) \cite{foot_linear}. Combined with $D_C \propto D_S$ and $\lambda_S=\sqrt{D_S \tau_S}$, this leads to a linear dependence between the spin-relaxation time and the momentum scattering time, which points to the Elliott-Yafet-type spin-relaxation mechanism \cite{PRB80_Jozsa2009, APS57_Fabian2007}. The dominance of this mechanism and therefore the spin-relaxation due to impurity scattering in our samples can be explained by impurity-induced spin-orbit coupling as described in Ref.~\citenum{PRL103_Castro2009}. Fig.~\ref{fig:Fig4}(b) shows $\lambda_S/D_S$ for different FLG samples \cite{foot_slopes}. Depending on whether the points were taken in the electron or the hole regime, we see different scalings of the slopes with an increasing number of layers. The lower values for $\lambda_S/D_S$ in the hole conduction regime indicate that there is a higher spin-flip probability for each scattering event. Overall, $\lambda_S/D_S$ increases with the number of layers, showing a reduced chance for spin flip with an increasing number of layers. This demonstrates the enhanced spin transport by using more than one layer graphene.

\section{\label{sec:Conclusions}Conclusions}
We have successfully produced lateral spin field-effect transistors (FET) on 3- to 20-layer graphene samples and measured the charge and spin transport properties of these devices. The reduced influence of external potentials, including the applied gate voltage, on the charge transport with an increasing number of layers has been explained by the distribution of the charges between the layers. The shape of the resistance curve can be modeled if we include screening effects. Further broadening and asymmetry of the peak shaped $R_S$ vs $V_g$ curve have been explained by inhomogeneous background doping of the flakes. Our conductivity per layer for 20-layer graphene stays a factor of $2.4$ below the value for bulk HOPG. This points to extrinsic scattering events at the bottom or the top of the flake limiting the transport in our devices. As those will be at least partly screened, impurities in the layers will also have limiting effects. A weak temperature dependence of the resistance (data not shown) also points to the fact that impurities and static scatterers are the main limiting factors.

The spin transport quantities $D_S$ and $\tau_S$ have been studied as a function of the induced charge carriers $n_g$ and of the number of layers. $\tau_S$ increases approximately linearly with the number of layers showing the expected enhancement of the spin lifetime due to the screening of the scattering potentials that has been modeled for the charge transport. The diffusion coefficients for spin ($D_S$) and charge ($D_C$) show a decrease from SLG to BLG and then stay approximately constant. This shows that the number of layers has only a weak influence on the diffusion, pointing to a weak coupling between the layers. 

The spin-relaxation length $\lambda_S$ is mainly enhanced for $n_g=0$. Therefore, we see improvement primarily for the spin transport in the regime around $\sigma_\mathrm{min}$. This is due to the intrinsic charge carriers of FLG, which mask the effect of induced charge carriers. We would like to mention at this point that we see no considerable temperature effect on $\lambda_S$ in FLG (data not shown). This points to a negligible effect of phonons on the spin-relaxation. The enhancement of $\lambda_S$ due to screening effects in FLG therefore enables the fabrication of improved spin transport devices.

Finally, we calculate $D_C$, using the DOS of FLG obtained by the zone-folding scheme, and compare the result with $D_S$. As we observe that $D_C~\propto~D_S$ and $\lambda_S~\propto~D_S$, it seems that the spin-relaxation in our FLG samples is mainly due to the Elliott-Yafet mechanism, which is also the case for SLG. As the linear dependence in our FLG spin transport measurements is based on an increase of both values by only a factor of $\sim1.5$, this result is not yet conclusive and requires further research. The theoretical expected dominance of the D'yakonov-Perel' spin scattering mechanism \cite{TEPJ148_Huertas-Hernando2007, PRL103_Huertas-Hernando2009} is probably only measurable in cleaner samples \cite{PRB82_Zhou2010} with higher diffusion coefficients and higher mobilities $\mu$. Therefore, measurements on high quality graphene spin valve devices have to be performed.

\textit{Note added.} Related results focusing on the comparison between spin-relaxation in SLG and BLG have recently been posted \cite{c_Han2010}.
\begin{acknowledgments}
We would like to acknowledge H.~T.~Jonkman, B.~Wolfs, J.~Holstein, and S.~Bakker for technical support and P.~J.~Zomer for critically reading the manuscript. This work was financed by NanoNed, the Zernike Institute for Advanced Materials and the Foundation for Fundamental Research on Matter (FOM).
\end{acknowledgments}


\begin{thebibliography}{42}
\bibitem[{foo({\natexlab{a}})}]{foot_17layers}
\bibinfo{note}{We do not consider the points for 17-layer graphene here or in other analysis as this sample showed, in general, different behavior. This could be due to enhanced contact from the sides of the graphene stack because of a possible different angle in the deposition step. We still show the data for the sake of completeness.}
\bibitem[{foo({\natexlab{b}})}]{foot_sigmamin}
\bibinfo{note}{Since electrical gating is not possible in bulk graphite, no gate-related resistance dependence can be observed. Hence, we use $\sigma_\mathrm{min}=\sigma$ for graphite.}
\bibitem[{foo({\natexlab{c}})}]{foot_60meV}
\bibinfo{note}{The $\mathrm{FWHM}\approx 60 ~\mathrm{meV}$ is here due to the
  fact that, for $D_C$ in FLG similar to $D_C$ in SLG, we see a non-physical peak
  at $V_0$ that is flattened by using this broadening.}
\bibitem[{foo({\natexlab{d}})}]{foot_linear}
\bibinfo{note}{Note that due to the small change of $D_S$ and $\lambda_S$ as a
  function of $n_g$ we see an increase of both values by only a factor of
  $\sim1.5$.}
\bibitem[{foo({\natexlab{e}})}]{foot_slopes}
\bibinfo{note}{As most samples are highly doped, we are only able to determine reliably the slope of $\lambda_S$ vs $D_S$ for just one of the two conduction regimes.}
\end{thebibliography}
\end{document}